\documentclass[11pt]{article}
\usepackage[english]{babel}
\usepackage{amsmath,amsthm}
\usepackage{amsfonts}
\theoremstyle{definition}

\theoremstyle{remark}

\numberwithin{equation}{section}
\begin{document}
\title{Turbulent Bubbly Channel Flow:\\ Role of Bubble Deformability}
\author{Sadegh Dabiri$^1$, Jiacai Lu$^2$, Gretar Tryggvason$^1$ \\
\\\vspace{6pt} $^1$University of Notre Dame, Notre Dame, IN 46556, USA \\ $^2$Worsecter Polytechnic Institute, Worcester, MA 01609, USA}
\maketitle
%% The abstract (in this file, and that submitted as text to arXiv) should include the exact phrase
%% "fluid dynamics video" or "fluid dynamics videos"
\begin{abstract}
This article describes the fluid dynamics video: "Turbulent Bubbly Channel Flow: Role of Bubble Deformability." The effect of bubble deformability of the flow rate of bubbly upflow in a turbulent vertical channel is examined using direct numerical simulations. A series of simulations with bubbles of decreasing deformability shows a transition from flow where deformable bubbles remain in the middle of the channel to a situation where nearly spherical bubbles slide along the walls.\end{abstract}
% main text
\section{Introduction}
To isolate the effect of bubble deformability we have carried out several simulations of many buoyant bubbles rising in a turbulent upflow in a vertical channel, where the only aspect that we change between the different runs is the surface tension. The computational domain is a rectangular channel bounded by two flat vertical walls. The channel is periodic in the spanwise and the streamwise direction. The flow is driven upward by a specified pressure gradient. The dimensionless numbers in the problem are Reynolds number, $Re=\frac{\rho uw}{\mu}$, Eotvos number, $Eo=\frac{\rho g^2 d}{\sigma}$, and density and viscosity ratios.
The pressure gradient and Eotvos number are specifies \emph{a priori} and the Reynolds number is calculated from the average velocity \emph{a posteriori}. Eotvos number is decreased from 4.5 to 0.9. The decrease in the deformability of bubbles reduces the flow rate and causes the Reynolds number to drop from 3750 to 1700.

\end{document}